\documentclass[twocolumn,twoside]{IEEEtran}
\usepackage{mathpazo}
\usepackage{times}

\usepackage{amsmath}
\usepackage{amsfonts}
\usepackage{latexsym}
\usepackage{amssymb}
\usepackage{cite}

\usepackage{upref}
\usepackage{theorem}
\usepackage{graphicx}
\usepackage{psfrag}
\usepackage{multirow}

\theoremstyle{plain}
\theorembodyfont{\normalfont\slshape}

\newtheorem{thm}{Theorem$\!$}
\newenvironment{theorem}
{\begin{thm}\hspace*{-1ex}{\bf.}}{\end{thm}}

\newtheorem{lem}[thm]{Lemma$\!$}
\newenvironment{lemma}{\begin{lem}\hspace*{-1ex}{\bf.}}{\end{lem}}

\newtheorem{prop}[thm]{Proposition$\!$}

\newtheorem{cor}[thm]{Corollary$\!$}
\newenvironment{corollary}{\begin{cor}\hspace*{-1ex}{\bf.}}{\end{cor}}

\newtheorem{defn}[thm]{Definition$\!$}
\newenvironment{definition}{\begin{defn}\hspace*{-1ex}{\bf.}}{\end{defn}}

\newtheorem{xmpl}[thm]{Example$\!$}
\newenvironment{example}{\begin{xmpl}\hspace*{-1ex}{\bf.}}{\hfill$\Box$\end{xmpl}}

\newtheorem{cnstr}{Construction$\!$}
\newenvironment{construction}{\begin{cnstr}\hspace*{-1ex}{\bf.}}{\end{cnstr}}

\newcounter{enumrom}
\renewcommand{\theenumrom}{(\roman{enumrom})}

\makeatletter
\renewcommand{\@endtheorem}{\endtrivlist}
\makeatother

\makeatletter
\renewcommand{\thefigure}{{\@arabic\c@figure}}
\renewcommand{\fnum@figure}{{\bf Figure\,\thefigure}}
\makeatother

\newcommand{\cD}{\mathcal{D}}

\newcommand{\cF}{\mathcal{F}}

\newcommand{\mathset}[1]{\left\{#1\right\}}
\newcommand{\abs}[1]{\left|#1\right|}
\newcommand{\ceilenv}[1]{\left\lceil #1 \right\rceil}
\newcommand{\floorenv}[1]{\left\lfloor #1 \right\rfloor}
\newcommand{\parenv}[1]{\left( #1 \right)}

\newcommand{\be}[1]{\begin{equation}\label{#1}}
\newcommand{\ee}{\end{equation}}

\renewcommand{\leq}{\leqslant}

\renewcommand{\geq}{\geqslant}

\renewcommand{\Bbb}{\mathbb}

\newcommand{\Cref}[1]{Co\-ro\-lla\-ry\,\ref{#1}}

\renewcommand{\Bbb}{\mathbb}

\newcommand{\N}{{\Bbb N}}
\newcommand{\R}{{\Bbb R}}
\newcommand{\Z}{{\Bbb Z}}

\DeclareMathAlphabet{\mathbfsl}{OT1}{cmr}{bx}{it}

\DeclareMathOperator{\wt}{wt}
\DeclareMathOperator{\lcm}{lcm}
\DeclareMathOperator{\per}{per}

\newcommand{\od}{\overline{d}}
\newcommand{\fgv}{f_{\mathrm{GV}}}

\outer\def\proclaim #1. #2\par{\medbreak
 \noindent{\bf#1.\enspace}{\sl#2\par}%
 \ifdim\lastskip<\medskipamount \removelastskip\penalty55\medskip\fi}

\mathchardef\inn="3232
\renewcommand{\in}{{\,\inn\,}}

\begin{document}

%----------------- The Title Declarations ------------------------------

\title{\Huge\bf Correcting Limited-Magnitude Errors in the Rank-Modulation Scheme}

\author{\large
Itzhak~Tamo and
Moshe~Schwartz,~\IEEEmembership{Member,~IEEE}
\thanks{Itzhak Tamo is with the Department
   of Electrical and Computer Engineering, Ben-Gurion University,
   Beer Sheva 84105, Israel
   (e-mail: tamo@ee.bgu.ac.il).}
\thanks{Moshe Schwartz is with the Department
   of Electrical and Computer Engineering, Ben-Gurion University,
   Beer Sheva 84105, Israel
   (e-mail: schwartz@ee.bgu.ac.il).}
}

% make the title area
\maketitle

\bibliographystyle{IEEEtranS}

\maketitle

%%%%%%%%%%%%%%%%%%%%%%%%%%%%%%%%%%%%%%%%%%%%%%%%%%%%%%%%%%%%%%%%%%%%%%
%%%%%%%%%%%%%%%%%%%%%%%%%%%%%%%%%%%%%%%%%%%%%%%%%%%%%%%%%%%%%%%%%%%%%%
%%%%%%%%%%%%%%%%%%%%%%%%%%%%%%%%%%%%%%%%%%%%%%%%%%%%%%%%%%%%%%%%%%%%%%

\begin{abstract}
We study error-correcting codes for permutations under the infinity norm,
motivated by a novel storage scheme for flash memories called 
\emph{rank modulation}.
In this scheme, a set of $n$ flash cells are combined to create a single
virtual multi-level cell. Information is stored in the permutation induced
by the cell charge levels. Spike errors, which are characterized by a 
limited-magnitude change in cell charge levels, correspond to a low-distance
change under the infinity norm.

We define codes protecting against spike errors, called
\emph{limited-magnitude rank-modulation codes} (LMRM codes),
and present several constructions for these codes,
some resulting in optimal codes. These codes admit simple recursive,
and sometimes direct, encoding and decoding procedures.

We also provide lower and upper bounds on the maximal size of LMRM codes
both in the general case, and in the case where the codes form a subgroup
of the symmetric group. In the asymptotic analysis, the codes we construct
out-perform the Gilbert-Varshamov-like bound estimate.
\end{abstract}


%%%%%%%%%%%%%%%%%%%%%%%%%%%%%%%%%%%%%%%%%%%%%%%%%%%%%%%%%%%%%%%%%%%%%%
%%%%%%%%%%%%%%%%%%%%%%%%%%%%%%%%%%%%%%%%%%%%%%%%%%%%%%%%%%%%%%%%%%%%%%
%%%%%%%%%%%%%%%%%%%%%%%%%%%%%%%%%%%%%%%%%%%%%%%%%%%%%%%%%%%%%%%%%%%%%%

\section{Introduction}
\label{sec:intro}

\IEEEPARstart{I}{n} the race to dominate non-volatile information-storage
devices,
flash memory is a prominent contender.
Flash memory is an electronic non-volatile memory
that uses floating-gate cells to store
information~\cite{CapGolOliZan99}.  While initially, flash memory cells
used to contains a single bit of information, in the standard 
multi-level flash-cell technology of today, every
cell has $q>2$ discrete states, $\mathset{0,1,\dots,q-1}$,
and therefore can store $\log_{2}q$ bits. The flash memory changes the
state of a cell by injecting (\emph{cell programming})
or removing (\emph{cell erasing}) charge into/from the cell.

Flash memories possess an inherent asymmetry: 
writing is more time- and energy-consuming than
reading~\cite{CapGolOliZan99}. The main reason behind this asymmetry
is the iterative
cell-programming procedure designed to avoid over-programming
\cite{BanSerHas05} (raising the cell's charge level above its target
level). While cells can be programmed individually,
only whole blocks (today, containing approximately $10^5$ cells,
see \cite{CapGolOliZan99}) can be erased to
the lowest state and then re-programmed. Since over-programming can
only be corrected by the block erasure, in practice a conservative
procedure is used for programming a cell, where charge is injected
into the cell over quite a few rounds~\cite{BanSerHas05}. After every
round, the charge level of the cell is measured and the next-round
injection is configured.  The charge level of the cell is made to
gradually approach the target state until it achieves the desired
accuracy. The iterative-programming approach is costly in time and
energy.

Another major concern for flash memory is data reliability. The stored
data can be corrupted due to charge leakage, a long-term factor that causes
the data retention problem. The data can also be affected by other
mechanisms, including read disturbance, write
disturbance~\cite{CapGolOliZan99}, etc.
Many of the error mechanisms
have an asymmetric property: they make the cells' charge
levels drift in one direction. (For example, charge leakage makes the
cell levels drift down.) Such a drift of cell charge levels causes
errors in aging devices.
The problem of data corruption is
further aggravated as the number of levels in multi-level cells increases,
since this reduces the safety margins for correct reading and writing.

To address these issues, the \emph{rank-modulation scheme} has been
recently suggested \cite{JiaMatSchBru09}.  By removing the need to
measure absolute cell-charge levels, the new scheme eliminates the
risk of cell over-programming, and reduces the effect of asymmetric
errors. In this scheme, a virtual cell that is composed of $n$ cells
with distinct charge levels, induces a permutation which is used to
represent the stored information. Each cell has a \emph{rank} which
indicates its relative position when ordering the cells according to
descending charge-level.  The ranks of the $n$ cells induce a
permutation of $\mathset{1,2,\dots,n}$.

When writing or reading the cell charge levels, we only need to
compare the charge levels \emph{between cells}.  Thus, the rank-modulation
scheme eliminates the need to use the absolute values of cell levels
to store information.  Since there is no risk of over-programming and
the cell charge levels can take continuous values, a substantially
less conservative cell programming method can be used and the writing
speed can be improved. In addition, asymmetric errors become less
serious, because when cell levels drift in the same direction, their
ranks are not affected as much as their absolute values. This way both
the writing speed and the data reliability can be improved.

While the rank-modulation scheme alleviates some of the problems associated
with current flash technology, the flash-memory channel remains noisy and
an error-control mechanism is required. In this work we consider an
error model which corresponds to spike errors. Such errors are
characterized by a limited-magnitude change in the charge level of cells,
and readily translates into a limited-magnitude change in the rank of,
possibly, \emph{all} cells in the stored permutation. This corresponds
to a bounded-distance change under the $\ell_\infty$-metric. We
call codes protecting against such errors \emph{limited-magnitude
rank-modulation codes}, or LMRM-codes.

A similar error model for flash
memory was considered not in the context of rank modulation in
\cite{CasSchBohBru07}, while a different error-model (charge-constrained
errors for rank modulation) was studied in \cite{JiaSchBru08}.
Codes over permutations are also referred to as \emph{permutation arrays}
and have been studied in the past under different metrics
\cite{ChaKur69,Bla74,BlaCohDez79,VinHaeWad00,DinFuKloWei02,FuKlo04,ColKloLin04}.
Specifically, permutation arrays under the $\ell_\infty$-metric
were considered in \cite{LinTsaTze08}.

The main contribution of this paper is a set of constructions and
bounds for such codes. The constructions presented are applicable for a
wide range of parameters, and admit simple decoding and encoding procedures.
We also present bounds on code parameters both for the general case, as well
as for the more restricted case of subgroup codes. Most notably, we present
an asymptotically-good family of codes, with non-vanishing normalized distance 
and rate, which exceed the Gilbert-Varshamov-like lower bound estimate.

It is important to note that, independently and concurrently,
Kl{\o}ve, Lin, Tsai, and Tzeng \cite{KloLinTsaTze09} describe
Construction \ref{con:simple} and its immediate generalization,
Construction \ref{con:directproduct}.
As the overall overlap is small, and since the two constructions lead
to our Construction \ref{con:semidirect}, which we show to produce an optimal
code, we bring these first two here for the sake of completeness.

The rest of the paper is organized as follows. In Section
\ref{sec:defs} we define the notation, and introduce the error-model as well
as the associated $\ell_\infty$-metric. We proceed in Section
\ref{sec:cons} and present the code
constructions and encoding/decoding algorithms.
In Section 
\ref{sec:bounds} we investigate general bounds on LMRM codes, code-anticode
bounds, and asymptotic-form bounds.
We conclude in Section \ref{sec:conclusion} with a summary of the results and
a short concluding remarks.

%%%%%%%%%%%%%%%%%%%%%%%%%%%%%%%%%%%%%%%%%%%%%%%%%%%%%%%%%%%%%%%%%%%%%%
%%%%%%%%%%%%%%%%%%%%%%%%%%%%%%%%%%%%%%%%%%%%%%%%%%%%%%%%%%%%%%%%%%%%%%
%%%%%%%%%%%%%%%%%%%%%%%%%%%%%%%%%%%%%%%%%%%%%%%%%%%%%%%%%%%%%%%%%%%%%%

\section{Definitions and Notations}
\label{sec:defs}

For any $m,n\in\N$, $m\leq n$, let $[m,n]$ denote the set
$\mathset{m,m+1,\dots,n}$, where we also denote by $[n]$ the set $[1,n]$.
Given any set $A$ of cardinality $n$, we denote by $S_A$ the set of all
permutations over the set $A$. By convention, we use $S_n$ to denote
the set $S_{[n]}$.

We will use both the vector notation for permutations $f \in S_n$,
where $f=[f_1,f_2,\dots,f_n]$ denotes the permutation
mapping $i\mapsto f_i=f(i)$ for all $i\in [n]$,
and the cycle notation, where
$f=(f_1,f_2,\dots,f_k)$ denotes the permutation mapping
$f_i\mapsto f_{i+1}$ for $i\in [k-1]$ as well as
$f_k\mapsto f_1$. Given two permutations $f,g \in S_n$,
the product $fg$ is a permutation mapping $i\mapsto f(g(i))$
for all $i\in [n]$.

Let us consider $n$ flash memory cells which we name $1,2,\dots,n$. The
charge level of each cell is denoted by $c_i\in\R$ for all $i\in[n]$.
In the
\emph{rank-modulation scheme} defined in \cite{JiaMatSchBru09}, the charge
levels of the cells induce a permutation in the following way: The
induced permutation (in vector notation) is $[f_1,f_2,\dots,f_n]$ iff
$c_{f_1}>c_{f_2}>\cdots > c_{f_n}$.

The rank-modulation scheme is defined by two functions: an encoding
function $E:Q\rightarrow S_n$, which takes a symbol from the input
alphabet $a\in Q$ and maps it to a permutation $f=E(a)\in S_n$, and
a decoding function $D:S_n\rightarrow Q$. Since no channel is devoid of noise,
a stored permutation $f=E(a)$ may be corrupted by any of a variety of
possible disturbance found in flash memory (see \cite{CapGolOliZan99}).
Assuming the changed version of $f$, denoted $f'$, is not too corrupted, 
we would like the decoding function to restore the original information
symbol, i.e., $D(f')=a$.

For a measure of the corruption of a stored permutation we may use any of
a variety of metrics over $S_n$ (see \cite{DezHua98}). Given a metric
over $S_n$, defined by a distance function
$d:S_n\times S_n\rightarrow\N\cup\mathset{0}$, an \emph{error-correcting code}
is a subset of $S_n$ with lower-bounded distance between distinct members.

In \cite{JiaSchBru08},
the Kendall-$\tau$ metric was used, where the distance between two permutations
is the number of adjacent transpositions required to transform one into the
other. This metric corresponds to a situation in which we can bound the
total difference in charge levels, and the error-correcting codes are
therefore named \emph{charge-constrained rank-modulation codes}.

In this work we consider a different type of common error -- a limited-magnitude
spike error. Suppose a permutation $f\in S_n$ was stored by setting the
charge levels of $n$ flash memory cells to $c_1,c_2,\dots,c_n$. We say a
single \emph{spike error of limited-magnitude $L$} has occurred in the $i$-th
cell if the corrupted charge level, $c'_i$, obeys $\abs{c_i-c'_i}\leq L$.
In general, we say spike errors of limited-magnitude $L$ have occurred
if the corrupted charge levels of all the cells, $c'_1,c'_2,\dots,c'_n$, obey
\[\max_{i\in [n]}\abs{c_i-c'_i}\leq L.\]

Let us denote by $f'$ the permutation induced by the cell charge levels
$c'_1,c'_2,\dots,c'_n$ under the rank-modulation scheme.
Under the plausible assumption that distinct charge levels are not
arbitrarily close (due to resolution constraints and quantization at the
reading mechanism), i.e., $\abs{c_i-c_j}\geq \ell$ for some positive constant
$\ell\in\R$ for all $i\neq j$, a spike error of limited-magnitude $L$ implies
a constant $d\in\N$ such that
\[\max_{i\in [n]}\abs{f^{-1}(i)-f'^{-1}(i)} < d. \]
Loosely speaking, an error of limited magnitude cannot change the \emph{rank}
of the cell $i$ (which is simply $f^{-1}(i)$) by $d$ or more positions.

We therefore find it suitable to use the $\ell_\infty$-metric over $S_n$
defined by the distance function
\[d_\infty(f,g)=\max_{i\in [n]}\abs{f(i)-g(i)},\]
for all $f,g\in S_n$. Since this will be the distance measure used
throughout the paper, we will usually omit the $\infty$ subscript.

\begin{definition}
A \emph{limited-magnitude rank-modulation code (LMRM-code)} with parameters
$(n,M,d)$, is a subset $C\subseteq S_n$ of cardinality $M$, such that
$d_\infty(f,g)\geq d$ for all $f,g\in C$, $f\neq g$. (We will sometimes
omit the parameter $M$.)
\end{definition}

We note that unlike the charge-constrained rank-modulation codes of
\cite{JiaSchBru08}, in which the codeword is stored in the permutation
induced by the charge levels of the cells, here the codeword is stored
in the \emph{inverse} of the permutation.

It may be the case that the code $C$ forms a subgroup of the symmetric
group $S_n$, which we will denote by $C\leq S_n$. We shall call such a
code a \emph{subgroup code}. Since groups offer a rich structure, we
will occasionally constrain ourselves to discuss subgroup codes.

%%%%%%%%%%%%%%%%%%%%%%%%%%%%%%%%%%%%%%%%%%%%%%%%%%%%%%%%%%%%%%%%%%%%%%
%%%%%%%%%%%%%%%%%%%%%%%%%%%%%%%%%%%%%%%%%%%%%%%%%%%%%%%%%%%%%%%%%%%%%%
%%%%%%%%%%%%%%%%%%%%%%%%%%%%%%%%%%%%%%%%%%%%%%%%%%%%%%%%%%%%%%%%%%%%%%

\section{Code Constructions}
\label{sec:cons}

In this section we describe three constructions for LMRM subgroup codes.
The first two were discovered independently and concurrently by
\cite{KloLinTsaTze09}.
We begin our constructions with the following, which bears a resemblance
to the unidirectional limited-magnitude codes described in
\cite{AhlAydKhaTol04}. This construction will turn out to be a simple case
of a more general construction given later.

\begin{construction}
\label{con:simple}
Given $n,d\in\N$ we construct
\[C=\mathset{ f\in S_n ~|~ f(i)\equiv i \pmod{d}}.\]
Alternatively, for every $i\in [d]$ let
\[A_i=(d\Z+i)\cap [n]=\mathset{j\in [n] ~|~ j\equiv i\pmod{d}},\]
and define $C$ to be the direct product of the symmetric groups over the
$A_i$'s,
\[C=S_{A_1}\times S_{A_2}\times \dots \times S_{A_{d}}.\] 

\end{construction}

\begin{theorem}
The code $C$ from Construction \ref{con:simple} is an $(n,M,d)$-LMRM code
with
\[M=\parenv{\ceilenv{n/d}!}^{n\bmod d}\parenv{\floorenv{n/d}!}^{d-(n\bmod d)}.\]
\end{theorem}

\begin{IEEEproof}
The length and size of the code are easily seen to be as claimed. All we
have to do now is show that the minimal distance of the code is indeed
$d$. Let $f,g\in C$ be two distinct codewords, and let $i\in [n]$ be
such that $f(i)\neq g(i)$.  Since $f(i)\equiv g(i)\pmod{d}$ it follows
that $\abs{f(i)-g(i)}\geq d$, and so $d(f,g)\geq d$.
\end{IEEEproof}

This construction allows a simple encoding procedure. To simplify the
presentation let us assume that $d$ divides $n$. The encoder takes as input
an integer $m\in [0,M-1]$ (where $M$ is the size of the code), 
e.g., by translating from a string of $\floorenv{\log_2 M}$ binary
input symbols. The number $M$ can then
be written in base $(n/d)!$,
that is
\[ M=\sum_{i=0}^{d-1}m_i\parenv{(n/d)!}^{i},\]
where $0\leq m_i \leq (n/d)!-1$. Finally, for every $i$ we
map the $i$-th digit, $m_i$, to $S_{A_{i+1}}$ using some function
\[\cF_i:\mathset{0,1,\dots,(n/d)!-1}\rightarrow S_{A_{i+1}}.\]
There are numerous efficiently-computable functions to satisfy $\cF_i$, such
as the factoradic representation (see \cite{Lai88,Leh60,MarStr07}),
as well others (see \cite{Knu98} and references therein).
Then, by using $\mathset{\cF_0,\cF_1,...,\cF_{d-1}}$ the resulting
encoding becomes
\[m\mapsto \cF_0(m_0)\times \cF_1(m_1)\times\dots\times
\cF_{d-1}(m_{d-1}).\]
  
A straightforward decoding procedure is also obtainable.
Let us assume that $f\in C$ was stored, where $C$ is an $(n,M,d)$-LMRM code
from Construction \ref{con:simple},
while the retrieved
permutation was $f'\in S_n$.  We further assume that the maximum magnitude of
errors  introduced by the channel is $\floorenv{(d-1)/2}$, i.e.,
$\abs{f(i)-f'(i)}\leq \floorenv{(d-1)/2}$ for all $i\in [n]$.

Since $C$ is a code of minimum distance $d$, there is a unique codeword $f^*$
at distance at most $\floorenv{(d-1)/2}$ from $f'$. Recovering this codeword
is simple and may be done independently for each of the coordinates:
For every coordinate $i\in [n]$, there is a
unique $f^*_i\in [n]$ such that $|f^*_i-f'(i)|\leq \frac{d-1}{2}$ and
$f^*_i\equiv i\pmod{d}$. The recovered permutation $f^*\in S_n$ is given by
$f^*(i)=f^*_i$. By definition, $f^*\in C$,
and by the algorithm presented we also
have $d(f^*,f')\leq \floorenv{(d-1)/2}$, hence $f^*=f$ which is the
original permutation which was stored.

Finding the original input message may be accomplished by decomposing
$f\in C$ into a product of permutations from $S_{A_i}$ and applying
$\cF_i^{-1}$ appropriately.

We now extend the direct-product approach and generalize the previous
construction. First we introduce a new notation. Given $f\in S_n$,
and a set $A\subseteq\N$ of size $n$, we denote by $f_A$ the same
permutation but over $A$. More formally, assuming
$A=\mathset{a_1,a_2,\dots,a_n}$, with $a_1<a_2<\dots<a_n$, we set
\[f_A=[a_{f(1)},a_{f(2)},\dots,a_{f(n)}].\]
Furthermore, given a set $C\subseteq S_n$, we define
\[C_A=\mathset{f_A ~|~ f\in C}.\]

\begin{construction}
\label{con:directproduct}
Let $n,k\in N$, and define the sets
\[A_i=(k\Z+i)\cap [n],\]
for all $i\in [k]$. Furthermore, for all $i\in[k]$ let $C^i$ be an
$(n_i,M_i,d_i)$-LMRM code, with $n_i=\abs{A_i}$. We construct the code
$C\subseteq S_n$,
\[C=C^1_{A_1}\times C^2_{A_2}\times \dots \times C^k_{A_k}.\]
\end{construction}

\begin{theorem}
The code $C$ from Construction \ref{con:directproduct} is an $(n,M,d)$-LMRM
with $M=\prod_{i=1}^k M_i$, and $d=\min_{i\in [k]}kd_i$.
(By convention, the distance of a code with one codeword is defined
as infinity.)
\end{theorem}

\begin{IEEEproof}
Again, the length and size of the code are easily verified. In addition,
given $f,g\in C$, $f\neq g$, it is easy to see that $f(i)-g(i)$ is a multiple
of $k$, for any $i\in [n]$, and so the distance of each of the constituent
codes is scaled by $k$, giving the desired result.
\end{IEEEproof}

Before describing the next construction we briefly observe some properties
which may be thought of as analogues to the case of linear subspace codes.
The metric defined by $d_\infty$ over $S_n$ is a right invariant metric
(see \cite{DezHua98}), i.e., for any $f,g,h\in S_n$,
\[d_\infty(f,g)=d_\infty(fh,gh).\]
We can then define the \emph{weight} of a permutation $f\in S_n$ as
\[\wt(f)=d_\infty(f,\iota),\]
where $\iota$ denotes the identity permutation. Thus,
for any $C\leq S_n$, an $(n,d)$-LMRM subgroup code, it follows that
\[d=\min_{f\in C, f\neq\iota}\wt(f).\]

For convenience, given a set $H\subseteq S_n$, we denote
\begin{align*}
d(H) &= \min_{f,g\in H, f\neq g}d(f,g) \\
\od(H) &= \max_{f,g\in H, f\neq g}d(f,g)
\end{align*}
Finally, we recall the following notation: For $H,K\subseteq S_n$ we denote
\[H^K=\mathset{h^k=khk^{-1} ~|~ h\in H,k\in K}.\]

\begin{construction}
\label{con:semidirect}
Let $H$ and $K$ be subgroups of $S_n$ such that $H^K=H$ and
$H\cap K=\mathset{\iota}$. We construct the code $C$ from the
following semi-direct group product,
\[C = H \rtimes K \cong HK = \mathset{ hk ~|~ h\in H, k\in K}.\]
\end{construction}

\begin{theorem}
\label{th:semidirect}
The code from Construction \ref{con:semidirect} is an $(n,M,d)$-LMRM subgroup
code with $M=\abs{H}\abs{K}$ and
\[d\geq \max\mathset{d(H)-\od(K), d(K)-\od(H)}.\]
\end{theorem}

\begin{IEEEproof}
It is well known (see for example \cite{Hal99}) that if $H^K=H$ and
$H\cap K=\mathset{\iota}$ then
$HK=KH\leq S_n$ and $\abs{HK}=\abs{H}\abs{K}$.
Given $h\in H$ and $k\in K$, where $hk\neq\iota$,
then from the triangle inequality
\begin{align*}
d(\iota,hk) & \geq d(\iota,k)-d(hk,k)= \wt(k)-\wt(h)\\
& \geq d(K) - \od(H).
\end{align*}
Interchanging $h$ and $k$ gives the other lower bound.
\end{IEEEproof}

The lower bound on the distance given in Theorem \ref{th:semidirect},
which we shall call the \emph{design distance}, is often not tight as
is shown in the following example.

\begin{example}
\label{ex:construction}
Let us construct an LMRM code of length $n=6$ and distance $d=3$.
According to construction \ref{con:simple}, the code $S_2\times S_2\times S_2$
is a $(6,8,3)$-LMRM code.

We can improve this by looking at the code
$C_3\leq S_3$ defined by
\[C_3=\mathset{[1,2,3],[2,3,1],[3,1,2]},\]
i.e., the cyclic group of size $3$, which is a $(3,3,2)$-LMRM code. By
Construction \ref{con:directproduct}, the code $C_3\times C_3$ is a
$(6,9,4)$-LMRM code, providing us a larger code than the previous one, with
a larger distance.

Finally, let us define $K\leq S_6$, a $(6,2,5)$-LMRM code, as
\[K=\mathset{[1,2,3,4,5,6],[6,5,4,3,2,1]}.\]
It may be verified that $H=C_3\times C_3$ and $K$ can be used with
Construction \ref{con:semidirect}, resulting in a $(6,18,3)$-LMRM code.
We note that while the design distance guaranteed by Theorem
\ref{th:semidirect} is just $1$, the resulting distance of the code
is actually $3$.
\end{example}

%%%%%%%%%%%%%%%%%%%%%%%%%%%%%%%%%%%%%%%%%%%%%%%%%%%%%%%%%%%%%%%%%%%%%%
%%%%%%%%%%%%%%%%%%%%%%%%%%%%%%%%%%%%%%%%%%%%%%%%%%%%%%%%%%%%%%%%%%%%%%
%%%%%%%%%%%%%%%%%%%%%%%%%%%%%%%%%%%%%%%%%%%%%%%%%%%%%%%%%%%%%%%%%%%%%%

\section{Bounds}
\label{sec:bounds}

\subsection{General Bounds}

The first two bounds we present are the obvious analogues of the
Gilbert-Varshamov bound, and the ball-packing bound (see, for example,
\cite{MacSlo78}). We first define the \emph{ball} of radius $r$ and
centered about $f\in S_n$ as the set,
\[B_{r,n}(f)=\mathset{ g\in S_n ~|~ d(f,g)\leq r }.\]
As mentioned before, the $\ell_\infty$ metric over $S_n$ is right invariant,
and so the size of a ball depends only on $r$ and $n$, and not on the
choice of center. We will therefore denote by $\abs{B_{r,n}}$ the size of a
ball of radius $r$ in $S_n$.

\begin{theorem}
\label{th:gv}
Let $n$, $M$, and $d$, be positive integers such that
$\abs{B_{d-1,n}}M\leq n!$. Then there exists an $(n,M,d)$-LMRM code.
\end{theorem}
\begin{IEEEproof}
Consider the following procedure: We start with the entire set $S_n$ with
all permutations unmarked, as well as an empty code $C$.
At each step we choose an unmarked permutation, $f$, add it to $C$, and mark
the ball $B_{d-1,n}(f)$. We stop when there remain no unmarked permutations.
The resulting code has minimal distance at least $d$,
and the number of iterations (which equals the size of the code) is at least
$n!/\abs{B_{d-1,n}}\geq M$.
\end{IEEEproof}

Next is a ball-packing bound, which was already mentioned in
\cite{LinTsaTze08}, and which we bring for completeness.
\begin{theorem}
\label{th:ball}
Let $C$ be an $(n,M,d)$-LMRM code. Then
\[\abs{B_{\floorenv{(d-1)/2},n}}M\leq n!.\]
\end{theorem}
\begin{IEEEproof}
Since $\ell_\infty$ over $S_n$ is a metric, and by the definition of an
$(n,M,d)$-LMRM code, the balls of radius $\floorenv{(d-1)/2}$ centered about
the codewords of $C$ are disjoint, proving the claim.
\end{IEEEproof}

We now proceed to present two upper bounds which are stronger, in general,
than the ball-packing bound of Theorem \ref{th:ball}. The first pertains to
subgroup codes, while the second is more general. Before starting, we
recall some well-known results from group theory (see \cite{Hal99}).

Let $G$ be a subgroup of $S_n$. For any $i\in [n]$, the \emph{orbit
of $i$ under the action of $G$} is defined as the set
\[i^G=\mathset{ g(i) ~|~ g\in G }.\]
The \emph{stabilizer of $i$ under the action of $G$} is defined as
\[ G_i= \mathset{ g\in G ~|~ g(i)=i },\]
and is a subgroup of $G$. Furthermore,
\begin{equation}
\label{eq:orbit}
\abs{G}=|i^G| \cdot |G_i|.
\end{equation}

\begin{theorem}
\label{th:groupub}
If $C$ is an $(n,M,d)$-LMRM subgroup code, then
\[ M \leq \frac{n!}{(d!)^{\floorenv{n/d}}(n\bmod d)!}.\]
\end{theorem}

\begin{IEEEproof}
For convenience, let us denote $r=n\bmod d$, and $k=\floorenv{n/d}$.
Let us now consider $C$ as it acts on the $d$-subsets of $[n]$.
By \eqref{eq:orbit} we get
\[M=\abs{C}=\abs{[1,d]^C}\cdot\abs{C_{[1,d]}}\leq \binom{n}{d}\abs{C_{[1,d]}},\]
where the last inequality follows from the fact that the orbit of $[1,d]$ under
$C$ contains at most all the $d$-subsets of $[n]$.
We can take another similar step and get
\begin{align*}
M & \leq \binom{n}{d}\abs{C_{[1,d]}} = \binom{n}{d}\abs{[d+1,2d]^{C_{[1,d]}}}
\abs{C_{[1,d],[d+1,2d]}}\\
& \leq \binom{n}{d}\binom{n-d}{d}\abs{C_{[1,d],[d+1,2d]}},
\end{align*}
where $[d+1,2d]^{C_{[1,d]}}$ denotes the orbit of $[d+1,2d]$ under the action
of $C_{[1,d]}$, i.e., the stabilizer of $[1,d]$ under $C$, while
$C_{[1,d],[d+1,2d]}$ denotes the subgroup of $C$ stabilizing both $[1,d]$
and $[d+1,2d]$.

Reiterating the argument above we reach
\[M \leq \prod_{i=0}^{k-1}\binom{n-di}{d}
\abs{C_{[1,d][d+1,2d],\dots,[(k-1)d+1,kd]}}.\]
It is now easy to see that
\[C_{[1,d],[d+1,2d],\dots,[(k-1)d+1,kd]}=\mathset{\iota},\]
or else the minimum distance $d$ of $C$ would be violated.
Thus,
\[M\leq \prod_{i=0}^{k-1}\binom{n-di}{d}=\frac{n!}{(d!)^k r!}.\]
\end{IEEEproof}

We can strengthen the upper bound of Theorem \ref{th:groupub}
by showing that codes attaining it
with equality must also satisfy certain divisibility conditions.

A group $G\leq S_n$ is said to be \emph{transitive} if for any $i,j\in [n]$
there is a permutation $f\in G$ such that $f(i)=j$. By
\eqref{eq:orbit}, the size of such a group $G$ must be divisible by $n$,
since the orbit of $i$ under the action of $G$ is $[n]$.

Extending this definition, we say a group $G\leq S_n$ is \emph{$k$-homogeneous}
if for any two $k$-sets $A,B\subseteq [n]$, there exists a permutation
$f\in G$ such that $f(A)=B$, where $f(A)=\mathset{f(a) ~|~ a\in A}$.
It then follows from \eqref{eq:orbit}, that
the size of such a group $G$ must be divisible by $\binom{n}{k}$.

The following theorem was given in \cite{Cam76}:
\begin{theorem}
\label{th:homog}
Let $G\leq S_n$ be a $k$-homogeneous finite group, where $2k\leq n+1$. Then
$G$ is also $(k-1)$-homogeneous.
\end{theorem}
Hence, for a $k$-homogeneous group $G\leq S_n$, $2k\leq n+1$, the size of
the group $G$ is divisible by
\[K_{n,k}=\lcm\mathset{\binom{n}{k},\binom{n}{k-1},\dots,\binom{n}{1}}.\]

\begin{theorem}
\label{th:div}
Let $C\leq S_n$ be an $(n,M,d)$-LMRM subgroup code attaining the upper
bound of Theorem \ref{th:groupub} with equality, i.e.,
\[ M = \frac{n!}{(d!)^{\floorenv{n/d}}(n\bmod d)!}.\]
Then
\[ \left.\lcm\mathset{ K_{n-id,d} ~\left|~ 0 \leq i\leq \frac{n-2d+1}{d}\right.}
~\right|~ M .\]
\end{theorem}
\begin{IEEEproof}
If we examine the proof of Theorem \ref{th:groupub}, for $C$ to attain
the upper bound we must have $\abs{[1,d]^C}=\binom{n}{d}$. Thus, for any
$d$-subset $A\subseteq [n]$, there exists a permutation $f_A\in C$ such that
$f_A([1,d])=A$. It now follows, that for any two $d$-subsets $A,B\subseteq [n]$,
we have that $f_Bf_A^{-1}(A)=B$, and $f_Bf_A^{-1}\in C$ since $C$ forms a
subgroup. Hence, $C$ is $d$-homogeneous.
If $2d\leq n+1$ then by Theorem \ref{th:homog}
we have $K_{n,d} ~|~ M$. 

Continuing in the same manner,
the group $C_{[1,d]}$ may be viewed as a permutation group over $[n-d]$
by deleting the elements of $[d]$ and relabeling the rest. Again,
we must have $\abs{[d+1,2d]^{C_{[1,d]}}}=\binom{n-d}{d}$ which means that
$C_{[1,d]}$ is also $d$-homogeneous. Again, if $2d\leq n-d+1$ then
$K_{n-d,d}$ divides $\abs{C_{[1,d]}}$, but $\abs{C_{1,d}}$ divides $\abs{C}$
since $C_{[1,d]}\leq C$. Reiterating the above arguments proves the claim.
\end{IEEEproof}

It is also important to notice that if an $(n,M,d)$-LMRM subgroup code
$C$ exists, then $M ~|~ n!$ since $C\leq S_n$.
\begin{example}
\label{ex:bound}
Continuing Example \ref{ex:construction} we would like to find an upper
bound to LMRM subgroup codes of length $n=6$ and minimum distance $d=3$.

We first substitute $n$ and $d$ in the ball-packing bound of Theorem
\ref{th:ball}. We get an upper bound (not only for subgroup codes) of
$\floorenv{6! / 13}=55$ since the size of a ball of radius $1$ in $S_6$
equals $13$.

Setting $n=6$ and $d=3$ in Theorem \ref{th:groupub} we get an upper bound
of size $6!/(3!)^2=20$. If a $(6,20,3)$-LMRM subgroup code exists, then
by Theorem \ref{th:div} its size must be divisible by its length (since it
must be $1$-homogeneous). However, $6$ does not divide $20$, and the next
candidate for an upper bound, $19$, does not divide $6!=720$.
Thus, the resulting
upper bound is $18$. This makes the $(6,18,3)$-LMRM subgroup code from
Example \ref{ex:construction} optimal.
\end{example}

\subsection{Codes and Anticodes}

We turn to describe another powerful bounding technique. The resulting
bounds bear a striking resemblance to the code-anticode method
of Delsarte \cite{Del73} and the set-antiset method of Deza \cite{DezFra77}.
However, both methods are not directly applicable to the case at hand. 

Given a metric space with integer distances, we can construct a graph
whose vertices are the points in the space, and an edge connects two vertices
if and only if they are at distance $1$ from each other. We call this
the \emph{induced graph} of the metric. If the metric distance between any two
points in the space equals the length of the shortest path between the corresponding
vertices in the induced graph (i.e., the distance in the graph),
we say the metric space is \emph{graphic}.

The code-anticode method of Delsarte
requires a graphic metric space which forms a
distance-regular graph. In our case, the $\ell_\infty$-metric over $S_n$
is not even graphic, and hence the code-anticode method does not apply.
The set-antiset method requires a metric over $S_n$ which is both
right and left invariant. Again, the $\ell_\infty$ metric-fails to
meet the method's requirements since it is not left invariant.

Given a set $A\subseteq S_n$, we denote
\[ \cD(A)=\mathset{ d(f,g) ~|~ f,g\in A }.\]
We also denote the \emph{inverse} of $A$ as
\[ A^{-1}=\mathset{ f^{-1} ~|~ f\in A}.\]

\begin{definition}
Two sets, $A,B\subseteq S_n$ are said to be \emph{a set and an antiset} if
\[ \cD(A)\cap \cD(B) = \mathset{0}.\]
\end{definition}

The following is the set-antiset bound for right-invariant metrics over $S_n$.

\begin{theorem}
\label{th:antiset}
Let $d:S_n\times S_n \rightarrow \N\cup\mathset{0}$ be a distance
measure inducing a right-invariant metric. Let $A,B\subseteq S_n$ be a
set and an antiset. Then
\[ \abs{A}\cdot\abs{B}\leq\abs{S_n}=n!.\] 
\end{theorem}

\begin{IEEEproof}
It is obvious that 
\[A^{-1}B=\mathset{f^{-1}g ~|~ f\in A, g\in B}\subseteq S_n.\]
We contend that $\abs{A^{-1}B}=\abs{A^{-1}}\cdot\abs{B}=\abs{A}\cdot\abs{B}$.
Let us assume the 
contrary, i.e., that there exist
$f_1,f_2\in A$ and $g_1,g_2\in B$ such that $f_1^{-1}g_1=f_2^{-1}g_2$ but
not both $f_1=f_2$ and $g_1=g_2$.

In that case, it follows that $g_1{g_2}^{-1}=f_1f_2^{-1} $.
We now have,
\[d(g_1,g_2)=d(g_1{g_2}^{-1},\iota)=d(f_1f_2^{-1},\iota)=
d(f_1,f_2).\]
But then
\[d(g_1,g_2)=d(f_1,f_2)\in\cD(A)\cap\cD(B)=\mathset{0}\]
implying that $g_1=g_2$ and $f_1=f_2$, a contradiction.
\end{IEEEproof}

To apply the set-antiset method to LMRM codes we need the following definition.
\begin{definition}
A \emph{limited-magnitude rank-modulation anticode (LMRM-anticode)}
with parameters $(n,M,d)$, is a subset $A\subseteq S_n$ of cardinality
$M$, such that $d_\infty(f,g)\leq d$ for all $f,g\in A$.
\end{definition}

\begin{theorem}
\label{th:anticode}
Let $C$ be an $(n,M_C,d)$-LMRM code, and let $A$ be an
$(n,M_A,d-1)$-LMRM anticode. Then $M_A M_C\leq n!$.
\end{theorem}
\begin{IEEEproof}
By the definition of a code and an anticode it is easily seen that
$\cD(A)\cap\cD(C)=\mathset{0}$. The claim is then a direct consequence
of Theorem \ref{th:antiset}.
\end{IEEEproof}

Theorem \ref{th:anticode} generalizes previous results. It may be easily
verified that a ball of radius $\floorenv{(d-1)/2}$ centered about the
identity permutation $\iota$ is an $(n,d-1)$-LMRM anticode.
Thus, the ball-packing bound of Theorem \ref{th:ball} is
a special case of Theorem \ref{th:anticode}.

The following is a generalization of Theorem \ref{th:groupub} to
LMRM codes which are not necessarily subgroups.

\begin{theorem}
\label{th:notgroupub}
If $C$ is an $(n,M,d)$-LMRM code, then
\[ M \leq \frac{n!}{(d!)^{\floorenv{n/d}}(n\bmod d)!}.\]
\end{theorem}

\begin{IEEEproof}
We construct the following $(n,M',d-1)$-LMRM anticode $A$: Let us denote
\[A_i=\parenv{[1,d]+(i-1)d}\cap [n].\]
We now define the anticode $A$ as
\[A=S_{A_1}\times S_{A_2}\times\dots\times S_{A_{\ceilenv{n/d}}}.\]
It is easy to verify that $A$ is indeed an anticode of maximum distance $d-1$,
and that its size is
\[M'=(d!)^{\floorenv{n/d}}(n\bmod d)!.\]
By Theorem \ref{th:anticode}, $M\cdot M'\leq n!$, and the claim on the maximal
size of an LMRM code follows.
\end{IEEEproof}

It should be noted that Theorem \ref{th:notgroupub} does not make
Theorem \ref{th:groupub} redundant, since through the proof of the latter
we were able to provide stricter necessary conditions for potential
subgroup codes attaining the bound with equality, as seen in
Theorem \ref{th:div}.

The next obvious question is: What is the size of the maximal size of an
$(n,d-1)$-LMRM anticode?

\begin{theorem}
\label{th:maxanti}
Let $A$ be an $(n,M,d-1)$-LMRM anticode. Then $M\leq (d!)^{n/d}$.
\end{theorem}
\begin{IEEEproof}
For all $1\leq i\leq n$ let $i^A=\mathset{f(i) ~|~ f\in A}$. It is
easy to see that $\abs{i^A} \leq d$, otherwise 
there would exist $f,g\in A$ such that $\abs{f(i)-g(i)}\geq d$ which
contradicts that maximal distance of $A$. 

Let $P$ be the following $n\times n$ binary matrix, where $P_{i,j}=1$ iff
there exists $f\in A$ such that $f(i)=j$, otherwise $P_{i,j}=0$.
It is well known (see for example \cite{Sch09}) that
\[\abs{A}\leq\per(P)=\sum_{f\in S_n}\prod_{i=1}^{n}P_{i,f(i)}\]
since all summands are either $0$ or $1$, and every permutation in $A$
corresponds to a non-vanishing summand.

According to Br\'{e}gman's Theorem (see \cite{Bre73}), for any
$n\times n$ binary matrix $P$ with $r_i$ $1$'s in the $i$-th row
\[\per(P)\leq \prod_{i=1}^n(r_i !)^{\frac{1}{r_i}}.\]
In our case, every row of $P$ contains at most $d$ $1$'s. We can certainly
change some $0$'s into $1$'s in $P$ so that every row contains \emph{exactly}
$d$ $1$'s, and by doing so, only increase the value of $\per(P)$.
It now follows that
\[M=\abs{A}\leq\per(P)\leq(d!)^{n/d}.\]
\end{IEEEproof}

Thus, for the case of $d|n$ we have an optimal anticode:

\begin{corollary}
The anticode constructed as part of Theorem \ref{th:notgroupub} is optimal
when $d|n$.
\end{corollary}

When $d$ does not divide $n$ the anticodes constructed in the proof of
Theorem \ref{th:notgroupub} are not necessarily optimal.  The
following theorem shows we can build larger anticodes.

\begin{theorem}
\label{th:notinvanti}
Let us denote $r=n\bmod d$. Then
there exists an $(n,M',d-1)$-LMRM anticode of size
\[
M'=\frac{(d!)^{\floorenv{\frac{n}{d}}-1}(d-r)!\floorenv{\frac{d+r}{2}}!
\ceilenv{\frac{d+r}{2}}!}
{\parenv{\floorenv{\frac{d+r}{2}}-r}!\parenv{\ceilenv{\frac{d+r}{2}}-r}!}.
\]
\end{theorem}
\begin{IEEEproof}
Consider the following $(d+r)\times (d+r)$ binary matrix $P$:
\newcommand{\pz}{\phantom{0}}
\newcommand{\po}{\phantom{1}}
\[
P=\begin{pmatrix}
\begin{array}{|ccc|}
\hline
\po & \po & \po \\
\po & 1_{\ceilenv{(d+r)/2}\times d} & \po \\
\po & \po & \po \\
\hline
\end{array}
\begin{array}{|c|}
\hline
\pz \\
0_{\ceilenv{(d+r)/2}\times r} \\
\pz \\
\hline
\end{array} \\
\begin{array}{|c|}
\hline
\pz \\
0_{\floorenv{(d+r)/2}\times r} \\
\pz \\
\hline
\end{array}
\begin{array}{|ccc|}
\hline
\po & \po & \po \\
\po & 1_{\floorenv{(d+r)/2}\times d} & \po \\
\po & \po & \po \\
\hline
\end{array}
\end{pmatrix},
\]
where $1_{i\times j}$ (respectively, $0_{i\times j}$) denotes the all $1$'s 
(respectively, all $0$'s) matrix of size $i\times j$.
It may now be verified that
\[\per(P)=(d-r)!\prod_{i=0}^{r-1}
\parenv{\floorenv{\frac{d+r}{2}}-i}\parenv{\ceilenv{\frac{d+r}{2}}-i}.\]

We now construct the following $n\times n$ binary matrix $Q$:
\[
Q=\parenv{\begin{array}{c@{}c@{}c@{}c@{}c}
\begin{array}{|c|}
\hline
1_{d\times d}\\
\hline
\end{array} & & & & \\
& \begin{array}{|c|}
\hline
1_{d\times d}\\
\hline
\end{array} & & \text{\Large 0} & \\
& & \ddots & & \\
& \text{\Large 0} & & \begin{array}{|c|}
\hline
1_{d\times d}\\
\hline
\end{array} & \\
& & & & 
\begin{array}{|c|}
\hline
P\\
\hline
\end{array}
\end{array}}
\]
where along the diagonal we have $\floorenv{\frac{n}{d}}-1$ blocks of
$1_{d\times d}$.

All the rows contain a contiguous block of $1$'s
of size $d$, and thus, all the permutations contributing to $\per(Q)$
form an anticode of maximum distance $d-1$. It can be easily seen that
\[M'=\per(Q)=(d!)^{\floorenv{\frac{n}{d}}-1}\per(P),\]
as claimed.
\end{IEEEproof}

With these anticodes we get the following two theorems.
\begin{theorem}
\label{th:goodanticode}
If $C$ is an $(n,M,d)$-LMRM code, then 
\[
M\leq \frac
{n!\parenv{\floorenv{\frac{d+r}{2}}-r}!\parenv{\ceilenv{\frac{d+r}{2}}-r}!}
{(d!)^{\floorenv{\frac{n}{d}}-1}(d-r)!\floorenv{\frac{d+r}{2}}!
\ceilenv{\frac{d+r}{2}}!},
\]
where $r=n\bmod d$.
\end{theorem}
\begin{IEEEproof}
Simply use the size of the anticodes of Theorem \ref{th:notinvanti} with
Theorem \ref{th:anticode}.
\end{IEEEproof}

\begin{theorem}
\label{th:almostn}
The optimal $(n,n-1)$-LMRM code, $n\geq 3$, has size $3$.
\end{theorem}
\begin{IEEEproof}
By Theorem \ref{th:goodanticode} we have
the following upper bound on the size of $(n,n-1)$-LMRM codes:
\[
\frac{n!}{(n-2)!\floorenv{\frac{n}{2}}\ceilenv{\frac{n}{2}}}=\begin{cases}
\frac{n-1}{n}\cdot 4 & \text{$n$ even} \\
\frac{n}{n+1}\cdot 4 & \text{$n$ odd}
\end{cases}
\]
and since the size must be an integer, it cannot exceed $3$. Such a code
can be easily constructed for any $n\geq 3$ and is simply the cyclic
group of order $3$ on the coordinates $\mathset{1,2,n}$:
\[C=\mathset{\iota, (1, 2, n), (1, n, 2)}\]
given in cycle notation.
\end{IEEEproof}

On a side note, Theorem \ref{th:almostn} was also
shown in \cite{KloLinTsaTze09}
using ad-hoc arguments.
Whether other infinite
families can be shown to be optimal using these anticodes is still unresolved.

\subsection{Asymptotic Bounds}

Some of the constructions and bounds presented in previous sections take on
a simple asymptotic form, which we explore below. We will compare the resulting
asymptotic bounds with those implied by the previous constructions of
\cite{LinTsaTze08}.

\begin{definition}
Given an $(n,M,d)$-LMRM code, we say it has \emph{rate}
$R=\frac{\log_2 M}{n}$ and \emph{normalized distance} $\delta=\frac{d}{n}$.
\end{definition}

A slight peculiarity arises here: One might expect the rate of a code
to be defined as $\frac{\log_2 M}{\log_2 n!}$ and not
$\frac{\log_2M}{\log_2 2^n}=\frac{\log_2 M}{n}$ since the ambient space
$S_n$ is of size $n!$. However, doing so results in asymptotic bounds
equal to $0$.

We begin with the asymptotic form of Theorem \ref{th:goodanticode}, and remind
that the binary entropy function $H_2: [0,1]\rightarrow [0,1]$
is defined as
\[H_2(p)=-p\log_2 p - (1-p)\log_2 (1-p).\]
\begin{theorem}
\label{th:aanti}
For any $(n,M,d)$-LMRM code,
\begin{align*}
R & \leq \parenv{\delta\floorenv{\frac{1}{\delta}}-\delta}\log_2
\parenv{\floorenv{\frac{1}{\delta}}-1}
+H_2\parenv{\delta\floorenv{\frac{1}{\delta}}-\delta}+\\
&\quad +2-2\delta\floorenv{\frac{1}{\delta}}+o(1).
\end{align*}
\end{theorem}
\begin{IEEEproof}
According to Theorem \ref{th:goodanticode},
\[
M\leq \frac
{n!\parenv{\floorenv{\frac{d+r}{2}}-r}!\parenv{\ceilenv{\frac{d+r}{2}}-r}!}
{(d!)^{\floorenv{\frac{n}{d}}-1}(d-r)!\floorenv{\frac{d+r}{2}}!
\ceilenv{\frac{d+r}{2}}!},
\]
where $r=n\bmod d$.
Moving to the $R$ and $\delta$ notation and slightly simplifying
the expression we get
\begin{align*}
2^{Rn} &\leq \frac{n!}{\parenv{(\delta n)!}^{\floorenv{\frac{1}{\delta}}-1}
\parenv{
\floorenv{\frac{n\parenv{\delta+1-\delta\floorenv{\frac{1}{\delta}}}}{2}}!
}^2} \cdot \\
&\quad
\cdot\frac{\parenv{\floorenv{\frac{
n\parenv{\delta-1+\delta\floorenv{\frac{1}{\delta}}}}{2}}!}^2}
{\parenv{n\parenv{\delta-1+\delta\floorenv{\frac{1}{\delta}}}}!}
\cdot 2^{o(n)}.
\end{align*}
At this point we use the well-known Stirling's approximation,
$m!=\sqrt{2\pi m}(m/e)^m(1+O(1/m))$. After rearranging we get
\[
2^{Rn}\leq \frac{
2^{\parenv{2-2\delta\floorenv{\frac{1}{\delta}}}n}}
{\delta^{\parenv{\delta\floorenv{\frac{1}{\delta}}-\delta}n}
\parenv{\delta+1-\delta\floorenv{\frac{1}{\delta}}}^{
\parenv{\delta+1-\delta\floorenv{\frac{1}{\delta}}}n}}
\cdot 2^{o(n)}.
\]
We take $\log_2$ of both sides, divide by $n$, and do some rearranging to reach
\begin{align*}
R & \leq 2 - 2\delta\floorenv{\frac{1}{\delta}}
-\parenv{\delta\floorenv{\frac{1}{\delta}}-\delta}\log_2\delta \\
& \quad -\parenv{\delta+1-\delta\floorenv{\frac{1}{\delta}}}
\log_2 \parenv{\delta+1-\delta\floorenv{\frac{1}{\delta}}} + o(1)\\
& = \parenv{\delta\floorenv{\frac{1}{\delta}}-\delta}\log_2
\parenv{\floorenv{\frac{1}{\delta}}-1}
+H_2\parenv{\delta\floorenv{\frac{1}{\delta}}-\delta}+\\
&\quad +2-2\delta\floorenv{\frac{1}{\delta}}+o(1)
\end{align*}
as claimed.
\end{IEEEproof}

For the next two asymptotic forms we need an estimate on the size of a ball
in the $\ell_\infty$-norm. While for any \emph{fixed} radius $r$, tight
asymptotic bounds on $\abs{B_{r,n}}$ are given in \cite{Sch09}, we require
an estimate for $r=\Theta(n)$. The best estimate,
to our knowledge,  for $0\leq r\leq \frac{n-1}{2}$, was given in \cite{Klo08}:
\begin{align}
\label{eq:ballsize}
\abs{B_{r,n}}&\geq \frac{\sqrt{2\pi n}}{2^{2r}}\parenv{\frac{2r+1}{e}}^n, \\
\label{eq:ballsizeup}
\abs{B_{r,n}}&\leq \parenv{(2r+1)!}^{\frac{n-2r}{2r+1}}
\prod_{i=r+1}^{2r}(i!)^{\frac{2}{i}}.
\end{align}
For our purposes, however, we do require an upper bound on $\abs{B_{r,n}}$ for
the entire range $0\leq r\leq n-1$. Therefore, we present an augmentation
of \eqref{eq:ballsizeup} in the following lemma.

\begin{lemma}
\label{lem:ballupper}
For all $0\leq r\leq n-1$,
\[
\abs{B_{r,n}}\leq \begin{cases}
\parenv{(2r+1)!}^{\frac{n-2r}{2r+1}}
\prod_{i=r+1}^{2r}(i!)^{\frac{2}{i}} & 0\leq r\leq \frac{n-1}{2}, \\
\parenv{n!}^{\frac{2r+2-n}{n}}
\prod_{i=r+1}^{n-1}(i!)^{\frac{2}{i}} & \frac{n-1}{2} \leq r \leq n-1.
\end{cases}
\]
\end{lemma}
\begin{IEEEproof}
It is easily seen that $B_{r,n}(\iota)$ is the set of all permutations
corresponding to non-vanishing terms in $\per(A)$ where $A$ is the
binary banded Toeplitz matrix defined by $A_{i,j}=1$ iff
$\abs{i-j}\leq r$. This observation has been used both in \cite{Sch09}
and in \cite{Klo08}.

The upper bound is immediately derived by using Br\'{e}gman's Theorem.
For example, for $\frac{n-1}{2}\leq r\leq n-1$, the matrix $A$ has
$2r+2-n$ rows with $n$ $1$'s, and two rows with $i$ $1$'s for each
$r+1\leq i\leq n-1$.
\end{IEEEproof}

We now state the asymptotic form of the Gilbert-Varshamov-like bound
of Theorem \ref{th:gv}.
\begin{theorem}
\label{th:agv}
For any constant $0 < \delta \leq 1$ there exists an infinite
sequence of $(n,M,d)$-LMRM codes with $\frac{d}{n}\geq \delta$ and
rate $R=\frac{\log_2 M}{n}$ satisfying $R\geq \fgv(\delta)+o(1)$, where
\[ \fgv(\delta)=\begin{cases}
\log_2\frac{1}{\delta}+2\delta(\log_2 e-1)-1 & 0 < \delta \leq \frac{1}{2}\\
-2\delta\log_2\frac{1}{\delta}+2(1-\delta)\log_2 e & \frac{1}{2}\leq \delta \leq 1
\end{cases}
\]
\end{theorem}
\begin{IEEEproof}
By Theorem \ref{th:gv} we are guaranteed the existence of an $(n,M,d)$-LMRM
code of size $M\geq n!/ \abs{B_{d-1,n}}$.
We can now use Lemma \ref{lem:ballupper} and replace $\abs{B_{d-1,n}}$ with
an appropriate upper bound.

Suppose $\frac{n-1}{2}\leq d-1\leq n-1$ (the
proof for the other case is similar). Then by Lemma \ref{lem:ballupper}
\begin{align*}
\abs{B_{\delta n-1,n}} & \leq 
(n!)^{2\delta-1}\prod_{i=\delta n}^{n-1}(i!)^{\frac{2}{i}} \\
& = \parenv{\frac{n}{e}}^{(2\delta-1)n}
\prod_{i=\delta n}^{n-1}\parenv{\frac{i}{e}}^2\cdot 2^{o(n)}\\
& = \frac{n^{(2\delta-1)n}}{e^n}
\parenv{\frac{(n-1)!}{(\delta n - 1)!}}^2\cdot 2^{o(n)}\\
& = \frac{n^n}{e^{(3-2\delta)n} \delta^{2\delta n}}\cdot 2^{o(n)},
\end{align*}
We now have
\[
2^{Rn} \geq \frac{n!}{\abs{B_{d-1,n}}}
\geq e^{(2-2\delta)n} \delta^{2\delta n}\cdot 2^{o(n)}.
\]
Taking $\log_2$ of both sides and dividing by $n$ completes the proof.
\end{IEEEproof}

\begin{figure*}[ht]
\psfrag{xax}{(a)}
\psfrag{xbx}{(b)}
\psfrag{xcx}{(c)}
\psfrag{xdx}{(d)}
\psfrag{xex}{(e)}
\psfrag{xfx}{(f)}
\psfrag{delta}{$\delta$}
\psfrag{rrr}{$R$}
\centering
\includegraphics[scale=1.0]{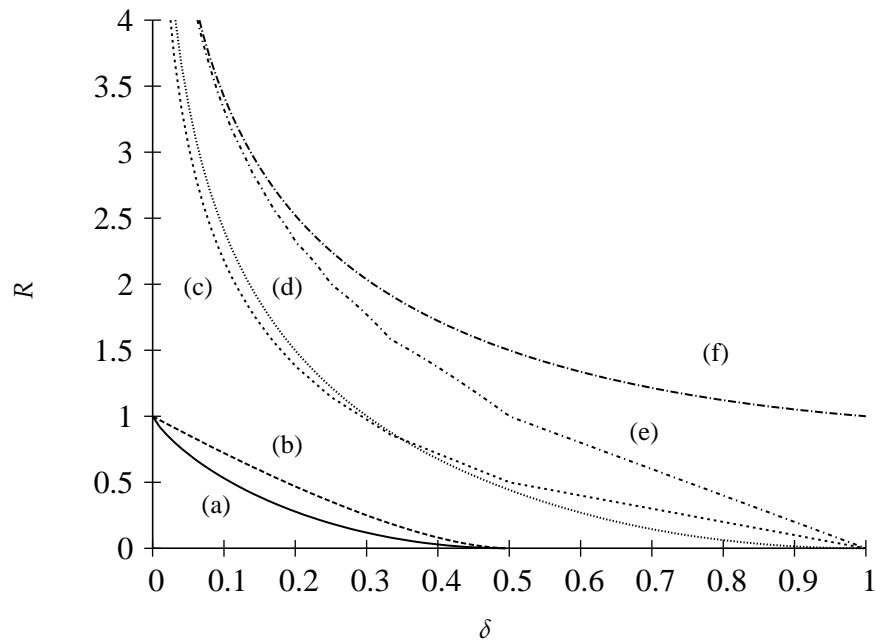}
\caption{
(a) The Gilbert-Varshamov bound in the $n$-cube
(b) The MRRW bound in the $n$-cube
(c) The rate of the code from Construction \ref{con:simple}
(d) The Gilbert-Varshamov-like bound of Theorem \ref{th:agv}
(e) The code-anticode bound of Theorem \ref{th:aanti}
(f) The ball-packing bound of Theorem \ref{th:aball}
}
\label{fig:bounds}
\end{figure*}

The ball-packing bound of Theorem \ref{th:ball} has the following
asymptotic equivalent:
\begin{theorem}
\label{th:aball}
For any $(n,M,d)$-LMRM code,
\[
R \leq \delta + \log_2\frac{1}{\delta} + o(1).
\]
\end{theorem}
\begin{IEEEproof}
The bound of Theorem \ref{th:ball} together with the lower bound of
\eqref{eq:ballsize} becomes,
\[
M \leq \frac{n!}{\abs{B_{\floorenv{(d-1)/2},n}}}
\leq \frac{n!2^{d'}}{\sqrt{2\pi n}}\parenv{\frac{e}{d'+1}}^n
\]
where $d'=d-2+(d\bmod 2)$.
Changing to the $R$ and $\delta$ notation, using Stirling's approximation,
and then taking $\log_2$ and dividing by $n$ gives as
\[R\leq \delta+\log_2\frac{1}{\delta}+o(1),\]
as desired.
\end{IEEEproof}

Finally, we analyze the asymptotics of the codes produced by
Construction \ref{con:simple}.
\begin{theorem}
\label{th:asimple}
For any constant $0 < \delta \leq 1$, Construction \ref{con:simple} produces
codes of rate
\begin{align*}
R & =\parenv{1-\delta\floorenv{\frac{1}{\delta}}}\log_2 
\parenv{\ceilenv{\frac{1}{\delta}}!} \\
& \quad + \parenv{\delta+\delta\floorenv{\frac{1}{\delta}}-1}\log_2
\parenv{\floorenv{\frac{1}{\delta}}!}.
\end{align*}
\end{theorem}
\begin{IEEEproof}
For any $(n,M,d)$-LMRM code produced by Construction \ref{con:simple} we
know that
\[M=\parenv{\ceilenv{n/d}!}^{n\bmod d}\parenv{\floorenv{n/d}!}^{d-(n\bmod d)}.\]
Just like before, we change to the $\delta$ and $R$ notation:
\[2^{Rn}=
\parenv{\ceilenv{\frac{1}{\delta}}!}^{
n\parenv{1-\delta\floorenv{\frac{1}{\delta}}}}
\parenv{\floorenv{\frac{1}{\delta}}!}^{
n\parenv{\delta+\delta\floorenv{\frac{1}{\delta}}-1}}.\]
We then take
$\log_2$ of both sides, and divide by $n$ to reach the claimed result.
\end{IEEEproof}

All the asymptotic bounds are shown in Figure \ref{fig:bounds}. Several
interesting observations can be made. First, the ball-packing bound 
of Theorem \ref{th:aball} is weaker than the code-anticode bound
of Theorem \ref{th:aanti}. This, however, may be due to a poor lower
bound on the size of a ball from \eqref{eq:ballsize}. It was conjectured
in \cite{Klo08} that this lower bound might be improved substantially.
We also note that Construction \ref{con:simple} produces codes which
asymptotically out-perform the Gilbert-Varshamov-like bound of
Theorem \ref{th:agv} for a wide range of $\delta$ (with crossover
at $\delta\approx 0.34904$), and appear to be
quite close to the bound otherwise. Again, this might be a result of a weak
upper bound on the size of a ball. Finally, the codes presented by
\cite{LinTsaTze08} are severely restricted since they are derived from
binary codes in the $n$-cube, and as such, are bounded by the $n$-cube
versions of the Gilbert-Varshamov bound and the MRRW bound
(see, for example, \cite{MacSlo78}).

%%%%%%%%%%%%%%%%%%%%%%%%%%%%%%%%%%%%%%%%%%%%%%%%%%%%%%%%%%%%%%%%%%%%%%
%%%%%%%%%%%%%%%%%%%%%%%%%%%%%%%%%%%%%%%%%%%%%%%%%%%%%%%%%%%%%%%%%%%%%%
%%%%%%%%%%%%%%%%%%%%%%%%%%%%%%%%%%%%%%%%%%%%%%%%%%%%%%%%%%%%%%%%%%%%%%

\section{Conclusion}
\label{sec:conclusion}

We have studied codes for the rank modulation scheme which protect against
limited-magnitude errors. We presented several code constructions which,
in some cases, produce optimal codes. The codes constructed can also be
encoded and decoded recursively, while the code of Construction
\ref{con:simple} may be encoded/decoded directly using a simple procedure
with small loss in rate. We note that all the constructions we presented
create codes which are subgroups of $S_n$.

We also explored bounds on the parameters of these codes. The strongest
upper bound appears to be the code-anticode bound of
Theorem \ref{th:notgroupub}. In the asymptotic study of these bounds, the
simple code from Construction \ref{con:simple} shows a better rate than
the one guaranteed by the Gilbert-Varshamov-like bound of
Theorem \ref{th:agv}, and the ball-packing upper bound of Theorem \ref{th:aball}
is always weaker than that of the code-anticode bound of Theorem \ref{th:aanti}.
Both, however, may be a result of a loose bound on the size of a ball in the
$\ell_\infty$-metric.

%%%%%%%%%%%%%%%%%%%%%%%%%%%%%%%%%%%%%%%%%%%%%%%%%%%%%%%%%%%%%%%%%%%%%%
%%%%%%%%%%%%%%%%%%%%%%%%%%%%%%%%%%%%%%%%%%%%%%%%%%%%%%%%%%%%%%%%%%%%%%
%%%%%%%%%%%%%%%%%%%%%%%%%%%%%%%%%%%%%%%%%%%%%%%%%%%%%%%%%%%%%%%%%%%%%%

\bibliography{allbib}
  
\end{document}